\documentclass{ifacconf}

\usepackage{natbib}       
\usepackage{amsmath, amssymb, amsfonts}
\usepackage{graphicx}
\usepackage{booktabs}
\usepackage{multirow}
\usepackage{mathrsfs}
\usepackage{float}
\usepackage{subfigure} 
\usepackage{enumerate}
\usepackage{verbatim}
\usepackage{color}
\usepackage{cases}
\usepackage{cancel}
\usepackage{hhline}
\usepackage{delarray}
\usepackage{url}

\begin{document}
\begin{frontmatter}

\title{Convex Model Predictive Control for Safe Output Consensus of Nonlinear Multi-Agent Systems} 

\thanks[footnoteinfo]{This work was supported by the Natural Science Foundation of China under Grant 62573014.}

\author[First]{Chao Wang} 
\author[Second]{Shuyuan Zhang} 
\author[First]{Lei Wang}

\address[First]{School of Automation Science and Electrical Engineering, Beihang University, Beijing, 100191, China (e-mail: wchao@buaa.edu.cn, lwang@buaa.edu.cn).}
\address[Second]{ICTEAM Institute, UCLouvain, 4 Avenue Georges Lema\^itre, 1348 Louvain-la-Neuve, Belgium (e-mail: shuyuan.zhang@uclouvain.be)}

\begin{abstract}               
Nonlinear dynamics and safety constraints typically result in a nonlinear programming problem when applying model predictive control to achieve safe output consensus. To avoid the heavy computational burden of solving a nonlinear programming problem directly, this paper proposes a novel Convex Model Predictive Control (CMPC) approach based on a Sequential Quadratic Programming (SQP) scheme. The core of our method lies in transforming the nonlinear constraints into linear forms: we linearize the system dynamics and convexify the discrete-time high-order control barrier functions using a proposed tangent-line projection method. Consequently, the original problem is reduced to a quadratic program that can be iteratively solved within the SQP scheme at each time step of CMPC. Furthermore, we provide the formal guarantee of the convergence of the SQP scheme, and subsequently guarantee the recursive feasibility and stability of CMPC. Simulations on multi-agent systems with unicycle dynamics demonstrate a 35–52 times reduction in computation time compared with baseline methods, confirming the suitability of the proposed approach for real-time safe output consensus control.
\end{abstract}


\end{frontmatter}
%==============================================================================
%% There are a number of predefined theorem-like environments in
%% ifacconf.cls:
%%
%% \begin{thm} ... \end{thm}            % Theorem
%% \begin{lem} ... \end{lem}            % Lemma
%% \begin{claim} ... \end{claim}        % Claim
%% \begin{conj} ... \end{conj}          % Conjecture
%% \begin{cor} ... \end{cor}            % Corollary
%% \begin{fact} ... \end{fact}          % Fact
%% \begin{hypo} ... \end{hypo}          % Hypothesis
%% \begin{prop} ... \end{prop}          % Proposition
%% \begin{crit} ... \end{crit}          % Criterion
%\theoremstyle{plain}
%\if@secthm
%\newtheorem{thm}{Theorem}[section]
%\@addtoreset{thm}{section}
%\else
%\newtheorem{thm}{Theorem}
%\fi
%\newtheorem{cor}[thm]{Corollary}
%\newtheorem{lem}[thm]{Lemma}
%\newtheorem{claim}[thm]{Claim}
%\newtheorem{axiom}[thm]{Axiom}
%\newtheorem{conj}[thm]{Conjecture}
%\newtheorem{fact}[thm]{Fact}
%\newtheorem{hypo}[thm]{Hypothesis}
%\newtheorem{assum}[thm]{Assumption}
%\newtheorem{prop}[thm]{Proposition}
%\newtheorem{crit}[thm]{Criterion}
%\theoremstyle{definition}
%\newtheorem{defn}[thm]{Definition}
%\newtheorem{exmp}[thm]{Example}
%\newtheorem{rem}[thm]{Remark}
%\newtheorem{prob}[thm]{Problem}
%\newtheorem{prin}[thm]{Principle}
%\newtheorem{alg}{Algorithm}

\section{INTRODUCTION}
\label{1}

The output consensus control of Multi-Agent Systems (MASs) has attracted significant research interest in recent years. It is widely used in fields such as multi-robot systems, sensor networks, and resource allocation (\cite{10383878, 9272828, WANG201519}). The objective of output consensus control is to design distributed controllers that synchronize outputs of all agents to a common trajectory or common value.

In practical applications, achieving consensus alone is insufficient; safety must be guaranteed simultaneously. This leads to the problem of safe output consensus, where agents must coordinate their behaviors while strictly adhering to physical constraints and collision avoidance requirements (\cite{11194736}).
For instance, in a swarm of unmanned aerial vehicles performing a search and rescue mission, agents must maintain a formation (consensus) while avoiding collisions with buildings, trees, and other drones. Consequently, incorporating safety constraints explicitly into the consensus control framework has become a critical research topic (\cite{10145590}).

To address safety specifications, the concept of a Control Barrier Function (CBF) was pioneered in (\cite{7040372}) to define a safe invariant set, ensuring that the system state remains within the safe region for all future time. The integration of CBFs with optimization-based control strategies has been extensively explored in the existing literature (\cite{MURAMATSU2024114, 7857061, 9777251, 9483029}).
A common approach involves incorporating CBFs into Quadratic Programming (QP) formulations. For instance, a QP-based method combined with CBFs was presented in (\cite{7857061}) to guarantee safety for multi-robot systems characterized by first-order dynamics. To accommodate systems with higher relative degrees, high-order control barrier functions were subsequently proposed, which are typically formulated within a QP framework to achieve safe control (\cite{9777251}). Despite their effectiveness in guaranteeing safety, QP-CBF methods rely exclusively on the current instantaneous state information, resulting in generating short-sighted control actions (\cite{9483029}).

To address this issue, several works have integrated CBF constraints into MPC. This combination strategically uses the long-horizon planning capabilities inherent in MPC while keeping the formal safety guarantees provided by CBFs. For coordinated multi-robot control, CBFs have been incorporated into distributed MPC schemes (\cite{10167791}). (\cite{OTSUKI20233138}) proposed a customized projected gradient descent approach to solve the resulting nonlinear optimization problem, showing faster computation speeds. For MASs with high relative degrees, a distributed MPC scheme using a Discrete-Time High-Order Control Barrier Function (DHCBF) was proposed in \cite{wang2025distri} to achieve safe consensus. While MPC-CBF methods generally show better control accuracy and performance compared to QP-CBF approaches, this method typically results in a non-convex Nonlinear Programming (NLP) problem, making the computation highly challenging for real-time applications.

Motivated by the non-convexity of existing approaches, this paper proposes a computationally efficient Convex Model Predictive Control (CMPC) approach for the safe output consensus of nonlinear MASs. Rather than solving the NLP directly, we adopt a Sequential Quadratic Programming (SQP) scheme, which converts the NLP within each MPC horizon into a sequence of convex QPs. Specifically, each QP subproblem is obtained by linearizing the nonlinear dynamics and convexifying the DHCBF constraints. Solving these QPs iteratively yields an efficient approximation of the optimal solution of the original problem.

The main contributions of this work are as follows:
\begin{enumerate}
\item We propose a novel CMPC approach for safe output consensus of nonlinear MASs based on an SQP scheme, in which a sequence of QP problems is solved iteratively. At each iteration, the nonlinear dynamics are linearized, and the DHCBF constraints are convexified via a proposed tangent-line projection method, rendering all constraints linear. Combined with the quadratic cost, this leads to a tractable QP at each time step.
    
\item We provide formal guarantees for the proposed method, including the local convergence of the SQP scheme, and the recursive feasibility and stability of the proposed CMPC approach, ensuring safe output consensus.
\end{enumerate}

The remainder of this paper is organized as follows. Section \ref{2} introduces the necessary preliminaries and formally defines the safe output consensus problem. The proposed CMPC approach is detailed in Section \ref{3}, with the corresponding theoretical analysis provided in Section \ref{4}. Section \ref{5} presents simulation results to validate the effectiveness of the proposed approach. Finally, Section \ref{6} concludes the paper.

\section{Preliminaries and Problem Formulation}
\label{2}

In this section, we first recall some fundamental definitions from DHCBF. Then, we introduce the safe output consensus problem.

\subsection{Discrete-Time High-Order Control Barrier Function}
\label{2.1}

Consider the discrete-time nonlinear system
\begin{equation}
    x(t+1) = f(x(t),u(t)),
    \label{eq:sys}
\end{equation}
where $t \in \mathbb{N}$ is the discrete time index, $x(t) \in \mathbb{R}^n$ is the system state, and $u(t) \in \mathbb{R}^m$ is the control input. The function $f: \mathbb{R}^n \times \mathbb{R}^m \to \mathbb{R}^n$ is a nonlinear function that describes the system dynamics.

Let the safe set $\mathcal{C} \subset \mathbb{R}^n$ be defined based on a continuous function $h: \mathbb{R}^n \to \mathbb{R}$, such that
\begin{equation}
    \mathcal{C} = \{x \in \mathbb{R}^n \mid h(x(t)) \ge 0\}.
    \label{eq:safe_set}
\end{equation}
A state $x(t)$ is considered safe if and only if $x(t) \in \mathcal{C}$.
The output $h(x(t))$ of system \eqref{eq:sys} is said to have relative degree $r$ if $h(f(x(t),u(t)))$ does not explicitly depend on $u(t)$ until the $r$-step forward difference.

To handle $h(x(t))$ with relative degree $r$ for system \eqref{eq:sys}, we define a sequence of discrete-time functions $h_{l}: \mathbb{R}^{n} \to \mathbb{R}$, $l=1,\ldots, r$ as:
\begin{equation}
h_{l}(x(t)) = \Delta h_{l-1}(x(t),u(t)) + \gamma_l h_{l-1}(x(t)),
\label{eq:hocbf-recursion}
\end{equation}
with $h_0(x(t))=h(x(t))$, where $\Delta h_{l-1}(x(t),u(t)) := h_{l-1}(f(x(t),u(t)))-h_{l-1}(x(t))$ and $\gamma_{l} \in (0,1]$ are tunable constants. Based on \eqref{eq:hocbf-recursion}, a corresponding sequence of safe sets is defined as
\begin{equation} \label{safesets}
\mathcal{C}_{l}:=\{ x \in \mathbb{R}^{n} \mid h_{l}(x(t)) \geq 0 \}, \ l=0,\ldots,r-1.
\end{equation}
Then, the DHCBF is defined as follows.
\begin{defn} [\cite{9777251}] \label{def1}
A continuous function $h: \mathbb{R}^{n} \to \mathbb{R} $ is a DHCBF with relative degree $r$ for system \eqref{eq:sys} if there exist $h_{r}(x(t))$ and $\mathcal{C}_{l}$, $l = 0, \ldots, r-1$, such that
\begin{equation}
h_r(x(t)) \ge 0, \ \forall x(t) \in \mathcal{C}_{0} \cap \dots \cap \mathcal{C}_{r-1}.
\label{eq:hocbf-condition}
\end{equation}
\end{defn}

The condition in \eqref{eq:hocbf-condition} guarantees the safety of system \eqref{eq:sys}.

\begin{lem} [\cite{9777251}] \label{lem1}
Given a DHCBF $h(x(t))$ from Definition \ref{def1} with safe sets defined by \eqref{safesets}, if $x(0) \in \mathcal{C}_{0} \cap \dots \cap \mathcal{C}_{l-1}$, then for $\forall t \geq 0$, any Lipschitz controller $u(t)$ that satisfies \eqref{eq:hocbf-condition} renders $\mathcal{C}_{0} \cap \dots \cap \mathcal{C}_{l-1}$ forward invariant for system \eqref{eq:sys}, i.e., $x(t) \in \mathcal{C}_{0} \cap \dots \cap \mathcal{C}_{l-1}$, $\forall t \geq 0$.
\end{lem}

\subsection{Graph theory}
\label{2.3}

The communication topology among $N$ agents is modeled by a directed graph  
$\mathcal{G} = (\mathcal{V}, \mathcal{E})$, where $\mathcal{V} = \{1,2,\dots,N\}$ denotes the set of nodes and 
$\mathcal{E} \subseteq \mathcal{V} \times \mathcal{V}$ denotes the set of directed edges. 
An edge $(j,i) \in \mathcal{E}$ indicates that agent $i$ can receive information from agent $j$. 
The in-neighbor set of agent $i$ is defined as
$\mathcal{N}_i = \{ j \in \mathcal{V} \mid (j,i) \in \mathcal{E} \}$.
The weighted adjacency matrix of $\mathcal{G}$ is denoted by $A = [a_{ij}]$, 
where $a_{ij} > 0$ if $(j,i) \in \mathcal{E}$ and $a_{ij}=0$ otherwise. 
The Laplacian matrix associated with $\mathcal{G}$ is defined as $L = D - A$, 
where $D = \mathrm{diag}(d_1,\dots,d_N)$ is the in-degree matrix with 
$d_i = \sum_{j \in \mathcal{N}_i} a_{ij}$. 
The directed graph $\mathcal{G}$ is said to contain a spanning tree if there exists at least one root node with a directed path to all other nodes.

\subsection{Problem Formulation}
\label{2.4}

Consider an MAS with $N$ agents, where the dynamics of each agent is described by
\begin{equation}\label{eq:dyn}
\begin{aligned}
x_i(t+1) &= f\big(x_i(t),u_i(t)\big),  \\
y_i(t+1)    &= C x_i(t+1), 
\end{aligned}
\end{equation}
for $i=1,\dots,N$, where $x_i(t)\in\mathbb{R}^n$ and $u_i(t)\in\mathbb{R}^m$ denote the state and control input of agent $i$ at time $t$, respectively, and $y_i(t)\in\mathbb{R}^p$ denotes the output with $C\in\mathbb{R}^{p\times n}$ the known output matrix. The nonlinear function $f:\mathbb{R}^n\times\mathbb{R}^m\to\mathbb{R}^n$ is assumed twice continuously differentiable. The aggregated vectors of all states and inputs in MASs \eqref{eq:dyn} at time step $t$ are defined as 
\begin{align}
X(t) &= [x_{1}^{\top}(t), \ldots, x_{N}^{\top}(t)]^{\top}, \\
U(t) &= [u_{1}^{\top}(t), \ldots, u_{N}^{\top}(t)]^{\top}.
\end{align}

Each agent is subject to bounded state and input constraints $x_i(t)\in\mathcal{X}$ and $u_i(t)\in\mathcal{U}$, where $\mathcal{X}\subset\mathbb{R}^n$ and $\mathcal{U}\subset\mathbb{R}^m$ are compact, convex sets containing the origin, ensuring that both the state and input remain within prescribed limits, i.e., $x_{\min} \leq x_i(t) \leq x_{\max}$, $u_{\min} \leq u_i(t) \leq u_{\max}$, where the inequalities are interpreted element-wise for given constant vectors $x_{\min}$, $x_{\max}$, $u_{\min}$ and $u_{\max}$.

This paper aims to achieve the consensus of output $y_{i}$ and ensure the safety of state $x_{i}$, which is called safe output consensus and described as follows.

\begin{prob}[Safe Output Consensus] \label{prb1}
Given a safe set $\mathcal{C} \subseteq \mathcal{X}$ like \eqref{eq:safe_set} for each agent, MAS \eqref{eq:dyn} is said to achieve the safe output consensus, if for $\forall \epsilon \geq 0$, there exists a constant $T \geq 0$ such that the following conditions hold for $\forall i \in \mathcal{N}$ and $j \in \mathcal{N}_{i}$:
\begin{align}
& \| y_{ij}(t) \| \leq \epsilon , t \geq	 T, \label{outcon} \\
& x_{i}(t) \in \mathcal{C}, t \geq 0,
\end{align}
where $y_{ij}(t) = y_{i}(t)-y_{j}(t)$.
\end{prob}

To solve the controller $u_{i}(t)$ for achieving the safe output consensus, we formulate Problem \ref{prb1} as the following optimization problem:
\begin{subequations}\label{optimization}
\begin{align}
 \min_{U} & \sum_{k=0}^{T_{p}-1} p(Y(k|t), U(k|t)) + q(Y(T_{p}|t)) \label{obj} \\
\text{s.t.} \quad & x_{i}(k+1|t) = f(x_{i}(k|t),u_{i}(k|t)), \label{opt:dyn} \\
\quad & y_{i}(k+1|t) = Cx_{i}(k+1|t), \label{opt:out} \\
\quad & x_{i}(0|t) = x_{i}(t), \label{opt:init} \\
\quad & u_{i}(k|t) \in \mathcal{U}, x_{i}(k|t) \in \mathcal{X}, \label{opt:cst} \\
\quad & x_{i}(k|t) \in \mathcal{C}, \label{safetycst}
\end{align}
\end{subequations}
where $ i \in \{1,\ldots,N\}, k \in \{0,\ldots,T_p-1\}$. In \eqref{obj}, $p(\cdot)$ and $q(\cdot)$ denote the stage and terminal costs. These cost functions are formulated to achieve the output consensus with the least control effort $u_i(k|t)$. The constraint \eqref{safetycst} ensures the state remains within the safe set $\mathcal{C}$, which encodes the obstacle avoidance and collision avoidance with other agents requirements.

The optimal solution to \eqref{optimization} at time $t$ are denoted as follows:
\begin{align}
\mathbf{U}^{\ast}(t) &= [U^{\ast, \top}(0|t), \ldots, U^{\ast, \top}(T_{p}-1|t)]^{\top}, \label{optimalu}
\end{align}
where $U^{\ast, \top}(0|t) = [u_{1}^{\ast, \top}(0|t), \ldots, u_{N}^{\ast, \top}(0|t)]$ is the optimal input sequence for system \eqref{eq:dyn} at time step $k=0$ prediceted at $t$. The first element of the optimal input sequence \eqref{optimalu}, $u_i^{\ast}(0|t)$, is applied to the system \eqref{eq:dyn}:
\begin{equation}\label{optsys}
\begin{aligned}
x_i(t+1) &= f(x_i(t), u_i^{\ast}(0|t) ), \\
y_i(t+1) &= C x_i(t+1), 
\end{aligned}
\end{equation}
to obtain the next state $x_{i}(t+1)$ and output $y_i(t+1)$. The constrained finite-time optimal control problem \eqref{optimization} is then re-solved at the next time step $t+1$ based on this new state.

Although our previous work (\cite{wang2025distri}) addressed problem \eqref{optimization} in a distributed manner, the resulting formulation remained a non-convex NLP, which limits computational efficiency. In this paper, we further reformulate the problem into a tractable convex optimization problem, thereby significantly improving the efficiency of the online computation.

\section{Convex Model Predictive Control}
\label{3}

In this section, we present our CMPC method. The optimization problem \eqref{optimization} is formulated as an SQP by linearizing the dynamics and convexifying the safety constraints.

\begin{figure}[t]
\centering
     \includegraphics[scale=0.28]{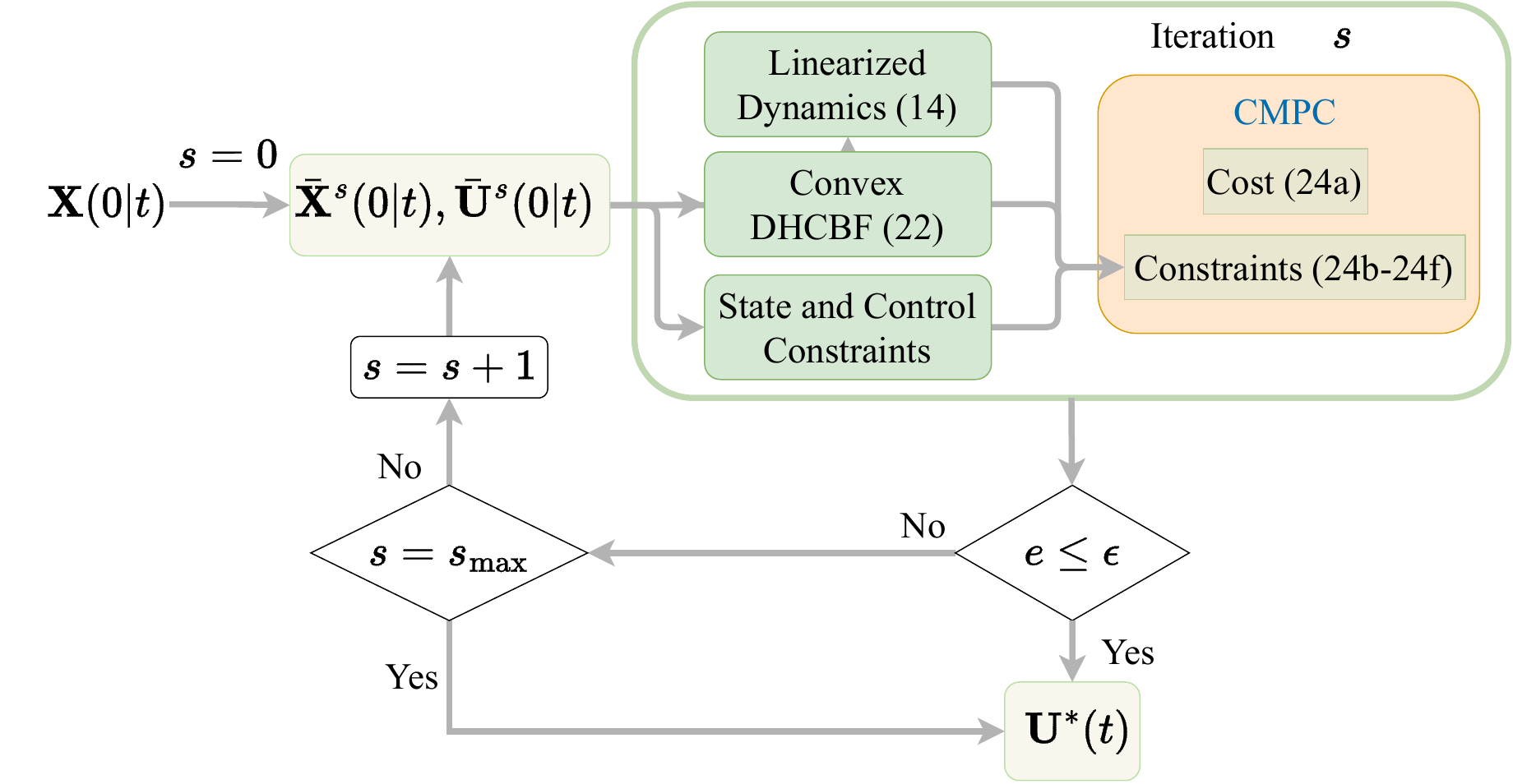}
 \caption{Structure of the iterative process at time step $t$.}
 \label{stru}
 \end{figure}

\subsection{Convex Optimization}
The structure of our method is described in Fig. \ref{stru}. At the current time step $t$, the optimization is initialized using the optimal results from the previous instant $t-1$. Specifically, the nominal states and inputs for iteration $s=0$ are set as $\bar{\mathbf{X}}(0|t) =\mathbf{X}^{\ast}(t-1)$ and $\bar{\mathbf{U}}(0|t)=\mathbf{U}^{\ast}(t-1)$. For the initial instance $t=0$, a zero-input guess $\bar{\mathbf{U}}^{0}(0) = 0$ is employed to generate the initial nominal state $\bar{\mathbf{X}}^{0}(0)$. Within each iteration $s$, a QP problem is formulated based on linearized dynamics and convex DHCBFs. Solving this problem yields the optimal states $\mathbf{X}^{\ast}(t)$ and inputs $\mathbf{U}^{\ast}(t)$, which are subsequently used to update the reference trajectories for the next iteration, given by $\bar{\mathbf{X}}^{s+1}(t) = \mathbf{X}^{\ast,s}(t)$ and $\bar{\mathbf{U}}^{s+1}(t) = \mathbf{U}^{\ast,s}(t)$. This iterative process terminates when the convergence error satisfies $e \leq \epsilon$ or the iteration count reaches the maximum $s_{\max}$. The optimized solution is forwarded to the CMPC formulation for the next time step, while the actual system state and output are updated via $x_{i}(t+1) = f(x_{i}(t), u_{i}^{\ast}(t))$ and $y_{i}(t+1)=Cx_{i}(t+1)$. The details in Fig. \ref{stru} are introduced in sections \ref{ld}-\ref{cm}.

\subsection{Linearization of Dynamics}
\label{ld}
At iteration $s$, the control input $\mathbf{u}_{i}^{s}(k|t)$ is refined by constructing a local linear model of the nonlinear dynamics around the nominal trajectory $\bar{x}_{i}^{s}(k|t)$,
\begin{align} 
& x_{i}^{s}(k+1|t)-\bar{x}_{i}^{s}(k+1|t) \notag \\ 
&\quad = A^{s}(x_{i}^{s}(k|t)-\bar{x}_{i}^{s}(k|t) )+B^{s}(u_{i}^{s}(k|t)-\bar{u}_{i}^{s}(k|t)). \label{linearsys} 
\end{align} 
The matrices $A^{s}$ and $B^{s}$ are obtained from the first-order expansion of the system dynamics $f(x_{i},u_{i})$:
\begin{align}
A^{s} &= \frac{\partial f}{\partial x_{i}}\Big|{(\bar{x}^{s}_{i}(k|t),\bar{u}^{s}_{i}(k|t))}, \label{As} \\
B^{s} & = \frac{\partial f}{\partial u_{i}}\Big|{(\bar{x}^{s}_{i}(k|t),\bar{u}^{s}_{i}(k|t))}. \label{Bs}
\end{align}
By updating the linear model iteratively, the nonlinear dynamics are successively approximated in a convex form.

\subsection{Convex Formulation of DHCBF}
\label{cf}
In this section, we show how to convexify the DHCBF into a linear formulation. 

\begin{figure}[t]
\centering
     \includegraphics[scale=0.26]{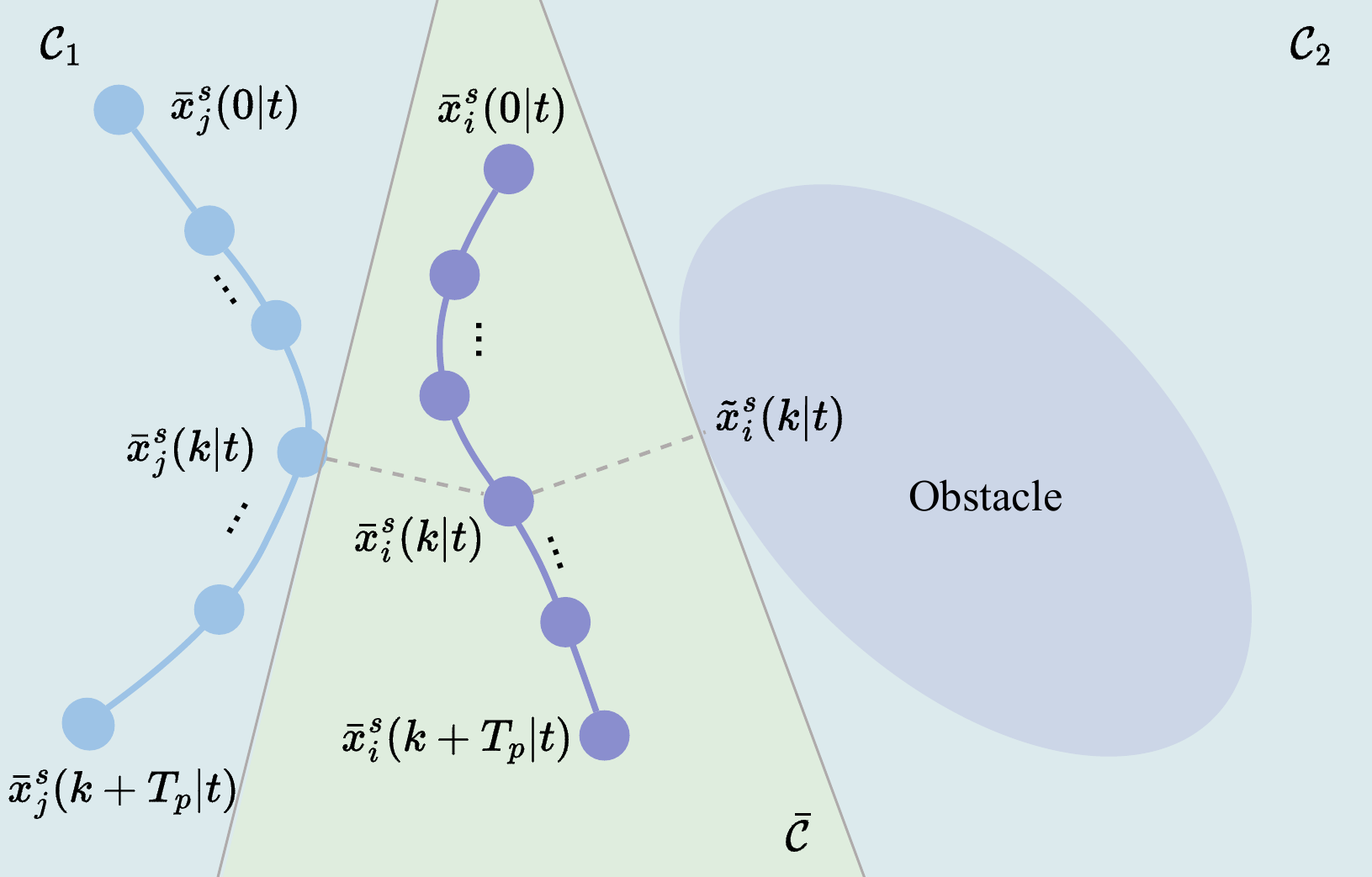}
 \caption{Convex formulation of safety constraints.}
 \label{cbf}
 \end{figure}

As illustrated in Fig. \ref{cbf}, at the time step $k$ predicted at $t$ of the $s$-th iteration, the safe set $\mathcal{C}$ is divided into three parts, including $\bar{\mathcal{C}}$ (green region), $\mathcal{C}_{1}$ and $\mathcal{C}_{2}$ (light blue region), via the tangent line projection. The safety constraints for inter-agent collision avoidance and obstacle avoidance are linearized by projecting two explicit tangent lines in the state space for the state $\bar{x}_{i}^{s}(k|t)$. The resulting linearized safety region $\bar{\mathcal{C}}$, defined as \eqref{barc}, is used in optimization. These lines intersect with the state of agent $j$ and the nearest point $\tilde{x}_{i}^{s}(k|t)$ on the obstacle boundary, respectively. It is worth noting that $\bar{x}_{i}^{s}(k|t)$ denotes the nominal state vector obtained from the previous iteration $(s-1)$, i.e., $\bar{x}_{i}^{s}(k|t) = x_{i}^{s-1}(k|t)$. The tangent lines passing through agent $j$ and the nearest obstacle point $\tilde{x}_{i}^{s}(k|t)$ are denoted by $h_{j}(\cdot)$ and $h_{o}(\cdot)$, respectively. Based on these definitions, the linearized safety constraints can be expressed as
\begin{align}
     h_{j}({x}_{i}^{s}(k|t), \bar{x}_{i}^{s}(k|t), \bar{x}_{j}^{s}(k|t)) \geq 0, \label{hj}\\
    h_{o}({x}_{i}^{s}(k|t), \bar{x}_{i}^{s}(k|t), \tilde{x}_{i}^{s}(k|t)) \geq 0. \label{ho}
\end{align}
For notational simplicity, $h_{j}(\cdot)$ and $h_{o}(\cdot)$ in \eqref{hj} and \eqref{ho} are denoted as $h_{j}({x}_{i}^{s}(k|t))$ and $h_{o}({x}_{i}^{s}(k|t))$, respectively, when necessary. Constraints \eqref{hj} and \eqref{ho} depict the resulting linearized safety region as follows:
\begin{equation}\label{barc}
\bar{\mathcal{C}}=\{x_{i}^{s}(k|t) \in \mathbb{R}^{n} \mid \bar{h}(x_{i}^{s}(k|t)) \geq 0\},
\end{equation}
where the function $\bar{h}(x_{i}^{s}(k|t)$ is not explicitly expressed, but it satisfies the following equality: 
\begin{align*}
\bar{h}(x_{i}^{s}(k|t)) \geq 0 
\Leftrightarrow
h_{j}(x_{i}^{s}(k|t)) \geq 0 
\cap 
h_{o}(x_{i}^{s}(k|t)) \geq 0
\end{align*}

The relative degrees of $h_{j}(\cdot)$ and $h_{o}(\cdot)$with respect to system \eqref{eq:dyn} are $r$. Based on Lemma \ref{lem1}, to guarantee safety with forward invariance, we derive a sequence of DHCBFs up to the order $r$:
\begin{align}\label{hstar}
h_{\star,l}(x_{i}^{s}(k|t)) &= \Delta h_{\star,l-1}(x_{i}^{s}(k|t)) \notag \\
& \quad + \gamma_{\star, l} h_{\star,l-1}(x_{i}^{s}(k|t)), 
\end{align}
where $h_{\star,0}(x_{i}^{s}(k|t))=h_{\star}(x_{i}^{s}(k|t))$, $\gamma_{\star, l} \in (0, 1]$, $l \in \{0,\ldots,r\}$ with $\star \in \{j,o\}$. Then, if there hold $h_{\star,l}(x_{i}^{s}(k|t)) \geq 0$ for $\forall t \geq 0$, the safety will be guaranteed.

Although the DHCBF constraints have been convexified into the formulation in \eqref{hstar}, the feasibility of the resulting optimization problem may be reduced due to the linearization and the sequence of difference operations. Inspired by (\cite{10156532}), we introduce a slack variable $w_{i,\star,l}^{s}(k|t) \in \mathbb{R}$ with a decay rate $(1-\gamma_{\star,l})$ to ensure feasibility. Consequently, the original constraints $h_{\star,l}(x_{i}^{s}(k|t)) \geq 0$ in \eqref{hstar} are relaxed as follows:
\begin{align}\label{slacklinear}
h_{\star,l-1}(x_{i}^{s}(k|t)) + \sum_{v=1}^{l}Z_{v,l}(1-\gamma_{\star,l})^{k}h_{\star,0}(x_{i}^{s}(v|t))& \notag \\
 \geq w_{i,\star,l}^{s}(k|t) Z_{0,l}(1-\gamma_{\star,l})^{k}h_{\star,0}(x_{i}^{s}(0|t))&,
\end{align}
where $l \in \{1,\ldots,r\}$. As the relative degree of $h_{\star}(\cdot)$ for system \eqref{eq:dyn} is $r$, the decision variable $u_{i}^{s}(k|t)$ appears explictly in $h_{\star, r-1}(x_{i}^{s}(k|t))$ as
\begin{align} 
h_{\star, r-1}(x_{i}^{s}(k|t)) & = a_{\star, r-1}^{s}(A^{s}(x_{i}^{s}(k|t)-\bar{x}_{i}^{s}(k|t) ) \notag \\
& \quad +B^{s}(u_{i}^{s}(k|t)-\bar{u}_{i}^{s}(k|t)))+b_{\star, r-1}^{s}, \label{hu}
\end{align}
where $a_{\star, r-1}^{s}, b_{\star, r-1}^{s} \in \mathbb{R}$ are computed using the tangent line projection method, and the coefficient of $u_{i}^{s}(k|t)$ is a real number $a^{s}B^{s}$. The coefficients $Z_{v,l}$ are constructed to maintain the linearity of \eqref{slacklinear} with respect to the decision variables. These coefficients are derived by recursively reformulating $h_{l-1}(\cdot)$ to $h_0(\cdot)$ over the horizon $\nu \in \{0,\dots,l\}$. Specifically, for $2 \le l$ and $\nu \le l-2$, $Z_{\nu,l}$ is defined as:
\begin{equation} \label{Z}
\begin{split}
    Z_{\nu,l} = \sum_{\mu=1}^{\mu_{\max}} [(\gamma_{\zeta_1} - 1)(\gamma_{\zeta_2} - 1) \cdots (\gamma_{\zeta_{l-\nu-1}} - 1)]_\mu, \\
    \zeta_1 < \zeta_2 < \cdots < \zeta_{l-\nu-1}, \zeta_\alpha \in \{1, 2, \dots, l-1\},
\end{split}
\end{equation}
where $[\cdot]_\mu$ represents the $\mu$-th combination product of the parenthesized terms, with the total number of combinations given by $\mu_{\max} = \binom{l-1}{l-\nu-1}$. The variable $\zeta_\alpha$ denotes the set of indices in (15). The boundary conditions are defined as follows: for $\nu = l-1$, we set $Z_{\nu,l} = -1$ if $l \ge 2$, and $Z_{\nu,l} = 1$ if $l=1$. Finally, for $\nu = l$, we define $Z_{\nu,l} = 0$.

\begin{rem}
The variable $\tilde{x}_{i}^{s}(k|t)$ in Fig. \ref{cbf} denotes the solution to the minimum-distance problem with respect to the safe set $\mathcal{C}_{2}$ computed at $\bar{x}_{i}^{s}(k|t)$. For smooth and differentiable DHCBFs, the mapping from $\bar{x}_{i}^{s}(k|t)$ to $\tilde{x}_{i}^{s}(k|t)$ can often be obtained explicitly (\cite{9867246}); for example, when $h(\cdot)$ is the $\ell_{2}$–norm and the obstacle is circular, $\tilde{x}_{i}^{s}(k|t)$ corresponds to the intersection of the segment from $\bar{x}_{i}^{s}(k|t)$ to the obstacle center with the boundary of $\mathcal{C}_{2}$. In more general cases such as elliptic constraints, an explicit form may not exist (\cite{skrypnik1994methods}), but $\tilde{x}_{i}^{s}(k|t)$ can be numerically approximated at each iteration using the known $\bar{x}_{i}^{s}(k|t)$ (\cite{10156532}).\end{rem}

\subsection{Convex Model Predictive Control}
\label{cm}
This section incorporates the linearized dynamics \eqref{linearsys} and the convexified safety constraints \eqref{slacklinear} into the optimization problem \eqref{optimization}, formulating the CMPC problem, which is solved at time step $t$ of iteration $s$ as follows:
\begin{subequations}\label{CMPC}
\begin{align}
& \min_{U^{s}, \Omega^{s}} \ 
 \sum_{k=0}^{T_{p}-1} p(Y^{s}(k|t),U^{s}(k|t),\Omega^{s}(k|t)) 
    + q(Y^{s}(T_{p}|t)) \label{cmpcobj} \\
\text{s.t. }
& x_{i}^{s}(k+1|t) - \bar{x}_{i}^{s}(k+1|t)= A^{s}\!\left(x_{i}^{s}(k|t)-\bar{x}_{i}^{s}(k|t)\right) \notag \\
&\qquad \qquad \qquad \qquad + B^{s}\!\left(u_{i}^{s}(k|t)-\bar{u}_{i}^{s}(k|t)\right), 
       \label{cmpcdyn} \\
& y_{i}^{s}(k+1|t) = C x^{s}_{i}(k+1|t), \label{cmpcout} \\
& x_{i}^{s}(0|t) = x^{s}_{i}(t), \label{cmpcinit} \\
& u_{i}^{s}(k|t) \in \mathcal{U},\
  x_{i}^{s}(k|t) \in \mathcal{X}, \label{cmpccst} \\
& h_{\star,l-1}\!\left(x_{i}^{s}(k|t)\right) 
  + \sum_{v=1}^{l} Z_{v,l}(1-\gamma_{\star,l})^{k}
    h_{\star,0}(x_{i}^{s}(v|t)) \notag \\
&\ge 
  w_{i,\star,l}^{s}(k|t)
  Z_{0,l}(1-\gamma_{\star,l})^{k}
  h_{\star,0}(x_{i}^{s}(0|t)). 
  \label{cmpcsafety}
\end{align}
\end{subequations}
where \eqref{cmpcobj} is the cost function with the stage cost and terminal cost as
\begin{align}
 & p(Y^{s}(k|t),U^{s}(k|t),\Omega^{s}(k|t)) = \sum_{i=1}^{N} (\sum_{j \in \mathcal{N}_{i}} \|y_{ij}^{s}(k|t)\|^{2}_{Q}  \notag \\
& \qquad \qquad+\|u_{i}^{s}(k|t)\|^{2}_{R} + \|w^{s}_{i, \star, l}(k|t)-1\|^{2}_{R_{w}}) , \label{convexp}\\
& q(Y^{s}(T_{p}|t)) = \sum_{i=1}^{N} \sum_{j \in \mathcal{N}_{i}} \|y_{ij}^{s}(T_{p}|t)\|^{2}_{P}, \label{convexq}
\end{align}
with $\Omega^{s}(k|t) = \{w^{s}_{1,\star,l}(k|t), \ldots, w^{s}_{N,\star,l}(k|t)\}$, $l \in \{1,\dots,r\}$, and weighting matrices $Q \succ 0$, $R \succ 0$, $R_{w} \succ 0$ and $P \succ 0$. The linearized dynamics in \eqref{linearsys} and the DHCBF constraints in \eqref{slacklinear} are enforced through \eqref{cmpcdyn} and \eqref{cmpcsafety} at each prediction step $k \in \{0,\ldots,T_{p}-1\}$. State and input constraints are incorporated via \eqref{cmpccst}. The slack variables are unconstrained, as feasibility is ensured by penalizing deviations from the nominal DHCBF conditions through the cost term $q(\cdot,\cdot,\Omega^{s}(k|t))$. To preserve the safety guarantees inherent to the DHCBFs, the constraints in \eqref{cmpccst} are strictly imposed for all $l \in \{1,\ldots,r\}$. 
The optimal decision variables at iteration $s$ consist of the control input sequence $\mathbf{U}^{\ast,s}(t)$ defined in \eqref{optimalu}, together with the slack-variable sequence $\mathbf{\Omega}^{\ast,s} = \{\Omega^{\ast,s}(0,t), \ldots, \Omega^{\ast,s}(T_{p}-1,t)\}$.

\begin{rem}
Note that the cost function \eqref{cmpcobj} becomes quadratic after applying \eqref{convexp} and \eqref{convexq}, and all constraints \eqref{cmpcdyn}--\eqref{cmpcsafety} are linear. Hence, the optimization problem \eqref{CMPC} is a QP. Since it is solved iteratively, the overall procedure corresponds to an SQP scheme, where each iteration involves a convex QP that can be solved efficiently. Moreover, to further enhance scalability, the method can be extended to a distributed implementation by adopting the neighbour-state estimation strategy proposed in our previous work (\cite{wang2025distri}).
\end{rem}

\section{Feasibility and Stability Analysis}
\label{4}
In this section, we first prove the convergence of the SQP at each time step, and then present the recursive feasibility and stability analysis of the proposed CMPC approach.

Consider the NLP formulated in \eqref{optimization} and the associated SQP \eqref{CMPC}. Denote the decision variable vector of \eqref{CMPC} as $z^s = [U^{s, \top}, X^{s, \top}, \Omega^{s, \top}]^\top$ at each iteration. Let $z^*$ be a local minimizer of \eqref{optimization} satisfying the Karush-Kuhn-Tucker (KKT) conditions. To prove the convergence of the SQP \eqref{CMPC}, we make the following assumption and lemma.

\begin{assum} [\cite{nocedal2006numerical}] \label{SQPcon}
Assume that the initial guess $z^0$ is sufficiently close to $z^*$.
\end{assum}

\begin{lem} [Theorem 18.3 in (\cite{nocedal2006numerical})] \label{lemsqp} 
If the Linear Independence Constraint Qualification (LICQ) holds at $z^*$ (i.e., the Jacobian matrix of active constraints has full row rank) and the Second-Order Sufficient Conditions (SOSC) are satisfied (i.e., the Hessian of the Lagrangian is positive definite on the null space of the active constraints), then the sequence $\{z^s\}$ generated by the SQP scheme converges locally to $z^*$.
\end{lem}

\begin{thm} \label{thmSQP}
If Assumption \ref{SQPcon} holds, then the sequence $\{z^s\}$ generated by solving \eqref{CMPC} converges locally to $z^*$.
\end{thm}

\textbf{Proof.}
The Jacobian matrix of the active constraints has full row rank due to its structure. Specifically, for the dynamics constraints \eqref{cmpcdyn}, the Jacobian with respect to the joint decision variables $(X^s, U^s)$ contains a square sub-matrix corresponding to the states $X^s$:
\[
\frac{\partial c_{\text{dyn}}}{\partial X^s} = 
\begin{bmatrix}
\mathcal{I}_{n} & 0 & \cdots & 0 \\
-A^s & \mathcal{I}_{n} & \cdots & 0 \\
\vdots & \ddots & \ddots & \vdots \\
0 & \cdots & -A^s & \mathcal{I}_{n}
\end{bmatrix}.
\]
The Jacobian matrix of the safety constraints with respect to the full decision vector $z^s = [{(U^s)}^\top, {(X^s)}^\top, {(\Omega^s)}^\top]^\top$ exhibits a specific block structure. 
Let $M$ denote the total number of scalar safety constraints defined in \eqref{cmpcsafety} across the prediction horizon $k=0, \dots, T_p-1$ and relative degrees $l=1, \dots, r$. 
Each row index $\iota \in \{1, \dots, M\}$ corresponds to a specific constraint instance defined by the tuple $(k, l, *)$, where $*$ represents the specific obstacle or agent index.
The Jacobian matrix is explicitly given by:
\[
J_{\text{safe}} = \frac{\partial c_{\text{safe}}}{\partial z^s} = 
\left[
\begin{array}{cc|cccc}
\frac{\partial h_{s}^{1}}{\partial U^s} & \frac{\partial h_s^{1}}{\partial X^s} & -C_{1} & 0 & \cdots & 0 \\
\frac{\partial h_s^{2}}{\partial U^s} & \frac{\partial h_s^{2}}{\partial X^s} & 0 & -C_{2} & \cdots & 0 \\
\vdots & \vdots & \vdots & \vdots & \ddots & \vdots \\
\frac{\partial h_s^{M}}{\partial U^s} & \frac{\partial h_s^{M}}{\partial X^s} & 0 & 0 & \cdots & -C_{M}
\end{array}
\right],
\]
where $h_{s}^{\iota}$ represents the constraint in \eqref{cmpcsafety}, $C_\iota$ is the coefficient of the slack variable for the $\iota$-th constraint, derived explicitly from the right-hand side of \eqref{slacklinear} as:
\begin{equation}
    C_\iota = Z_{0,l}(1-\gamma_{*,l})^{k}h_{*,0}(x_{i}^{s}(0|t)).
\end{equation}
The rightmost block of $J_{\text{safe}}$ is a diagonal matrix corresponding to the unique slack variables $\Omega^s \in \mathbb{R}^M$. Provided that the initial safety margin is non-zero (i.e., $h_{*,0}(x_{i}^{s}(0|t)) \neq 0$), we have $C_\iota \neq 0$, which ensures that this diagonal block has full rank. Consequently, the entire Jacobian matrix $J_{\text{safe}}$ satisfies the LICQ. Thus, LICQ is strictly satisfied. Furthermore, the Hessian matrix $H_{QP}$ of the cost function \eqref{cmpcobj} is a block-diagonal matrix with respect to the decision components $U^s, X^s, \Omega^s$:
\begin{equation}
    H_{QP} = \text{diag}(\mathbf{H}_U, \mathbf{H}_X, \mathbf{H}_\Omega).
\end{equation}
where $\mathbf{H}_U = I_{T_p} \otimes (I_N \otimes 2R)$ and $\mathbf{H}_\Omega = I_{M} \otimes 2R_{w}$ with $\otimes$ denoting the Kroneker product. Since $R, R_{w} \succ 0$, they are strictly positive definite. The state Hessian $\mathbf{H}_X = \text{diag}(\Phi_Q, \dots, \Phi_P)$ is positive semi-definite. This is because its sub-blocks, defined by $\Phi_Q = (L + L^\top) \otimes (C^\top Q C)$ and $\Phi_P = (L + L^\top) \otimes (C^\top P C)$, depend on the Laplacian $L$, which has a zero eigenvalue associated with the consensus subspace. However, the strict positive definiteness of $\mathbf{H}_U$, coupled with the system dynamics, ensures that the reduced Hessian projected onto the subspace defined by the linear dynamics is positive definite, thereby satisfying the SOSC condition. 
By Lemma \ref{lemsqp}, the sequence $\{z^s\}$ converges locally to the optimizer $z^*$. $\hfill \square$

\begin{rem}
In practice, it is usually difficult to find an initial guess $z^{0}$ as discussed in Assumption \ref{SQPcon}. Thus, when we apply our proposed method, the CMPC \eqref{CMPC} is solved iteratively at each time step $t$ and the solution is extracted once the convergence criteria or the maximum iteration number $s_{\max}$ is reached. The convergence criteria are designed as follows:
\begin{equation}\label{criteria}
\begin{aligned}
e_{abs}(\mathbf{Y}^{\ast,s}(t)) & = \| \mathbf{Y}^{\ast,s}(t)-\bar{\mathbf{Y}}^{\ast,s}(t) \|,  \\
e_{rel}(\mathbf{Y}^{\ast,s}(t)) & = \| \mathbf{Y}^{\ast,s}(t)-\bar{\mathbf{Y}}^{\ast,s}(t) \| / \|\bar{\mathbf{Y}}^{\ast,s}(t) \|.
\end{aligned}
\end{equation}
where $\mathbf{Y}^{\ast, s}(t) = [Y^{\ast, s, \top}(0|t), \ldots, Y^{\ast,s, \top}(T_{p}-1|t)]^{\top}$, $e_{abs}$ and $e_{rel}$ are absolute and relative errors, respectively. The iterative optimization stops when $e_{abs} \leq \epsilon_{abs}$ or $e_{rel} \leq \epsilon_{rel}$ with $\epsilon_{abs} >0$ and $\epsilon_{rel} > 0$. 
\end{rem}

Assuming that Theorem \ref{thmSQP} holds, the recursive feasibility of CMPC is guaranteed by Theorem \ref{thm:feasibility}. First, we denote $\eta \triangleq a B$ as the coefficient vector of $u_{i}$ in \eqref{hu}, where the superscript $s$ is omitted for notational simplicity, and let
\begin{align}
U_{\min} &= \inf_{u_{i} \in \mathcal{U}} [\eta u_i]  = \eta^- u_{\max} + \eta^+ u_{\min}, \notag \\
U_{\max} & = \sup_{u_{i} \in \mathcal{U}} [\eta u_i] = \eta^+ u_{\max} + \eta^- u_{\min}, \notag \\
 F_{\max} & = \sup_{x_{i} \in \mathcal{X}} [\sum_{v=1}^{r}-Z_{v,r}(1-\gamma_{\star,r})^{k}h_{\star,0}(x_{i}(v|t)) \notag \\
 & \quad +  Z_{0,r}(1-\gamma_{\star,l})^{k}h_{\star,0}(x_{i}(0|t))], \notag
\end{align}
where $\eta^+ = \max(\eta, 0)$ and $\eta^- = \min(\eta, 0)$ represent the element-wise maximum and minimum operations, respectively. Then, we introduce the following assumption, which is often used in the filed of MPC.

\begin{assum} [\cite{MAYNE2000789}] \label{ass:terminal}
For $\forall i \in \mathcal{V}$ and $j \in \mathcal{N}_{i}$, there exists control law $u_{i}(t) = \kappa_f(x_{i}(t)
, x_{j}(t)) \in \mathcal{U}$ such that for all $y_{ij}(t)\in \mathcal{Y}_{f}$, the terminal cost $q(\cdot)$ satisfies:
    \begin{equation} \label{ter}
        q(Y(t+1)) - q(Y(t)) + p(Y(t), U(t), \Omega(t)) \le 0, 
    \end{equation}
where $\mathcal{Y}_{f} \triangleq \{ y_{ij} \mid q(y_{ij}) \leq c) $ for some $c >0$, is the terminal invariant set.
\end{assum}

%\begin{assum} [\cite{704994}] \label{ass:slack}
%The slack variables $w_{i,*,l}^s$ introduced in \eqref{slacklinear} ensure that the feasible set is non-empty for any state prediction. The penalty weights in \eqref{convexp} are chosen sufficiently large to enforce $w \to 1$ when physically feasible.
%\end{assum}

\begin{thm} \label{thm:feasibility}
If the CMPC program \eqref{CMPC} is feasible at the initial time $t=0$ and $U_{\min} \geq F_{\max}$, then it remains feasible for all subsequent time steps $t > 0$, provided that Assumptions \ref{SQPcon} and \ref{ass:terminal} hold.
\end{thm}

\textbf{Proof.}
The proof proceeds by induction. Assume that the CMPC program is feasible at time $t$ and yields the optimal solution sequence $\mathbf{U}^{\ast}(t)$. The actual control input applied to system \eqref{eq:dyn} is $u_{i}(t) = u_{i}^{*}(0|t)$, resulting in the true state $x_{i}(t+1) = f(x_{i}(t), u_{i}^{*}(0|t))$ and the output $y_{i}(t+1) = C x_{i}(t+1)$.
At time $t+1$, we construct a candidate input sequence $\tilde{U}(t+1)$ based on the converged optimal solution:
\begin{equation*}
    \tilde{u}_{i}(k|t+1) = 
    \begin{cases} 
    u_{i}^{*}(k+1|t), & k = 0, \dots, T_p-2 \\
    \kappa_f(x_{i}^{*}(T_p|t), x_{j}^{*}(T_p|t)), & k = T_p-1
    \end{cases}
\end{equation*}
For recursive feasibility to hold, $\tilde{U}(t+1)$ must be feasible starting from the actual state $x_{i}(t+1)$.
Input constraints $\tilde{u}_{i} \in \mathcal{U}$ is satisfied by inheritance since $u^{*}_{i}(k|t) \in \mathcal{U}$ and $\kappa_f(\cdot) \in \mathcal{U}$ from Assumption \ref{ass:terminal}.  
The control bound on $u_{i}$ always satisfies
\begin{equation}
U_{\min} \leq aBu_{i} \leq U_{\max}.
\end{equation}
Since $U_{\min} \geq F_{\max}$, the intersection of the control bounds and linearized safety constraint  \eqref{cmpcsafety} is always nonempty. We conclude that there exist $u_{i} (t) \in \mathcal{U}$ such that $x_{i}(t+1) \in \mathcal{C}$ for all $t > 0$. Besides, as $\mathcal{C} \in \mathcal{X}$, there always holds $x_{i}(t+1) \in \mathcal{X}$. Thus, a feasible solution $(\tilde{U}, \tilde{\Omega})$ exists for the CMPC program at time $t+1$, with $\tilde{\Omega} = \mathcal{I}_{M}$.
$\hfill \square$

Next, we show the stability of CMPC \eqref{CMPC}, which implies the achievement of safe output consensus. We introduce the following assumption.

\begin{assum} \label{ass:sqp_convergence}
(\cite{nocedal2006numerical}). 
At each iteration, a trust-region control strategy is used, ensuring that the linearization error $\Delta_{lin} = \| f(x,u) - (A^s x + B^s u) \|$ remains bounded and is dominated by the decrease in the cost function.
\end{assum}

\begin{thm} \label{stability}
Given positive definite weighting matrices $Q, R, R_{w}, P$ satisfying \eqref{ter}, under Assumptions \ref{SQPcon}, \ref{ass:terminal} and \ref{ass:sqp_convergence}, the system \eqref{eq:dyn} achieves the safe output consensus by the CMPC program \eqref{CMPC}, if $U_{\min} \geq F_{\max}$.
\end{thm}

\textbf{Proof.}
Denote the value of the cost function \eqref{cmpcobj} at time step $t$ as $J(t)$, the optimal value as  $J^*(t)$. Consider $J(\cdot)$ as a Lyapunov candidate. Let $\tilde{J}(t+1)$ be the cost associated with the feasible candidate solution constructed in Theorem \ref{thm:feasibility}. By optimality, $J^*(t+1) \le \tilde{J}(t+1)$. The cost difference is:
\begin{align}
    & J^*(t+1)  - J^*(t) \le \tilde{J}(t+1) - J^*(t) \notag \\
    & = \sum_{i=1}^{N} \Delta_{lin}^{i}- p(0|t) \notag \\
    &+\underbrace{q(T_p|t+1) - q(T_p-1|t+1) + p(T_{p}-1|t+1)}_{\le 0 \text{ via \eqref{ter} in Assumption \ref{ass:terminal}}}
\end{align}
where $\Delta_{lin}^{i}$ represents the residual error due to the linearization dynamics \eqref{linearsys}.
According to Assumption \ref{ass:sqp_convergence}, the trust-region mechanism ensures that the cost reduction from the stage cost $p(0|t)$ dominates the linearization error $\Delta_{lin}$. Specifically:
\begin{equation}
    J^*(t+1) - J^*(t) \le - \alpha p(0|t)
\end{equation}
for some constant $\alpha \in (0, 1]$. Since $p(\cdot)$ is positive definite, $J^*(t)$ is a strictly decreasing sequence. Therefore, $J^*(t) \to 0$ as $t \to \infty$, which implies $y_{ij}(t) \to 0$ and $w(t) \to 1$. This proves that the output consensus is achieved asymptotically. Besides, as $U_{\min} \geq F_{\max}$, we conclude that the system remains safe for all $t \geq 0$, by Theorem \ref{thm:feasibility}. $\hfill \square$
 
 \section{Simulation}
 \label{5}
In this section, a numerical simulation is performed on a unicycle model to validate the effectiveness of the proposed CMPC approaches. Furthermore, its performance is quantitatively compared with other methods.
 
Consider a discrete-time multi-unicycle system consisting of a virtual leader, indexed as $0$, and three followers, indexed as $1$–$3$. The dynamics of each agent is described as follows:
\begin{equation}\label{simdyn}
\begin{aligned}
&
\begin{bmatrix}
p_{i,x}(t{+}1) \\
p_{i,y}(t{+}1) \\
\theta_{i}(t{+}1) \\
v_{i}(t{+}1)
\end{bmatrix}
=
\begin{bmatrix}
p_{i,x}(t)+v_{i}(t)\cos(\theta_{i}(t))\Delta t \\
p_{i,y}(t)+v_{i}(t)\sin(\theta_{i}(t))\Delta t \\
\theta_{i}(t)+u_{i,1}(t)\Delta t \\
v_{i}(t)+u_{i,2}(t)\Delta t
\end{bmatrix}, \\
& y_i(t{+}1)
=
\begin{bmatrix}
1 & 0 & 0 & 0 \\
0 & 0 & 1 & 0 \\
0 & 0 & 0 & 1
\end{bmatrix}
x_i(t{+}1)
=
\begin{bmatrix}
p_{i,x}(t{+}1) \\
\theta_{i}(t{+}1) \\
v_i(t{+}1)
\end{bmatrix},
\end{aligned}
\end{equation}
where $(p_{i,x}, p_{i,y})$ represent the position of agent $i$, while $\theta_{i}$ and $v_{i}$ denote its heading angle and linear velocity, respectively. The state and control input of agent $i$ are denoted as $x_{i} = [p_{i,x}, p_{i,y}, \theta_{i}, v_{i}]^{\top}$ and $u_{i} = [u_{i,1}, u_{i,2}]^{\top}$, respectively. The system output is denoted as $y_{i}$. The system is discretized with $\Delta t =0.1$. System \eqref{simdyn} is subject to the following state and input constraints:
\begin{align}
\mathcal{X} & = \{x_{i} \in \mathbb{R}^{n}: -10 \cdot \mathcal{I}_{4\times1} \leq x_{i} \leq 10 \cdot \mathcal{I}_{4 \times 1}\}, \\
\mathcal{U} & = \{u_{i} \in \mathbb{R}^{n}: [-0.3, -2]^{\top} \leq u_{i} \leq [0.3, 2]^{\top}\}.
\end{align}
The communication among agents is represented by a directed graph, whose Laplacian matrix is given as $L=
[0, 0, 0,0; -1, 2, -1, 0; 0, -1, 1, 0; 0, 0, -1, 1]$.
The initial states of all agents are $x_{1}=[-6.2, -0.5, 0, 0]^{\top}$, $x_{2}=[-5.6, 0, 0, 0]^{\top}$, $x_{3}=[-5, 0.5, 0, 0]^{\top}$ and $x_{4}=[-6.3, 1, 0, 0]^{\top}$. The goal state of the virtual leader is defined as $[5,0.3,0,0]^{\top}$. A circular obstacle is centered at $(-2,0)$ with $r_{o}=1$. Two candidate DHCBFs are defined as $h^{1}(x_{i}, x_{j}) = \|x_{i}-x_{j}\|^{2}-d^{2}$, and $h^{2}(x_{i}, x_{o}) = \|x_{i}-x_{o}\|^{2}-r_{o}^{2}$ to ensure safe of agent $i$, where $d=0.1$ is the prescribed safe distance. The resulting linearized DHCBFs by \eqref{hj} and \eqref{ho} are calculated as
\begin{align}
h_{j}(x_{i})  = & (\tilde{p}_{i,x}-p_{j,x})p_{i,x}+(\tilde{p}_{i,y}-p_{j,y})p_{i,y} \notag \\
& -(d^{2}-p_{j,x}^{2}-p_{j,y}^{2}+\tilde{p}_{i,x}p_{j,x}+\tilde{p}_{i,y}p_{j,y}) \label{simhj} \\
h_{o}(x_{i})  = & (\tilde{p}_{i,x}-p_{o,x})p_{i,x}+(\tilde{p}_{i,y}-p_{o,y})p_{i,y} \notag \\
& -(r_{o}^{2}-p_{o,x}^{2}-p_{o,y}^{2}+\tilde{p}_{i,x}p_{o,x}+\tilde{p}_{i,y}p_{o,y}) \label{simho} 
\end{align}
By \eqref{Z}, there are $Z_{0,2}=\gamma_{1}-1$, $Z_{1,2}=-1$, $Z_{0,1}=1$, $Z_{1,1}=Z_{2,2}=0$. The cost function is defined as \eqref{convexp} and \eqref{convexq}, where $Q=10 \cdot \mathcal{I}_{3}$, $R = 100 \cdot \mathcal{I}_{2}$, $R_{w} = 1000$ and $P = 500 \cdot \mathcal{I}_{3}$. The convergence errors in \eqref{criteria} are $\epsilon_{abs} = 10^{-4}$ and $\epsilon_{rel} = 10^{-2}$, and the maximum iteration number is $s_{\max}=30$.

\begin{figure}[t]
\centering
\subfigure[]{
		\includegraphics[scale=0.22]{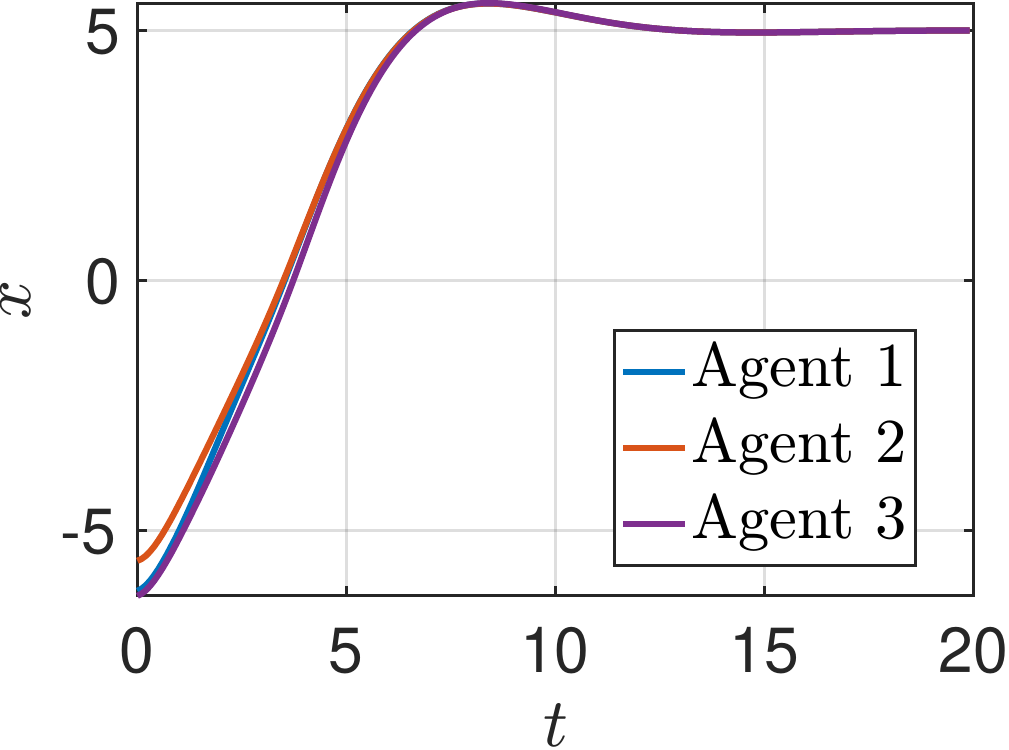}}
\subfigure[]{
		\includegraphics[scale=0.22]{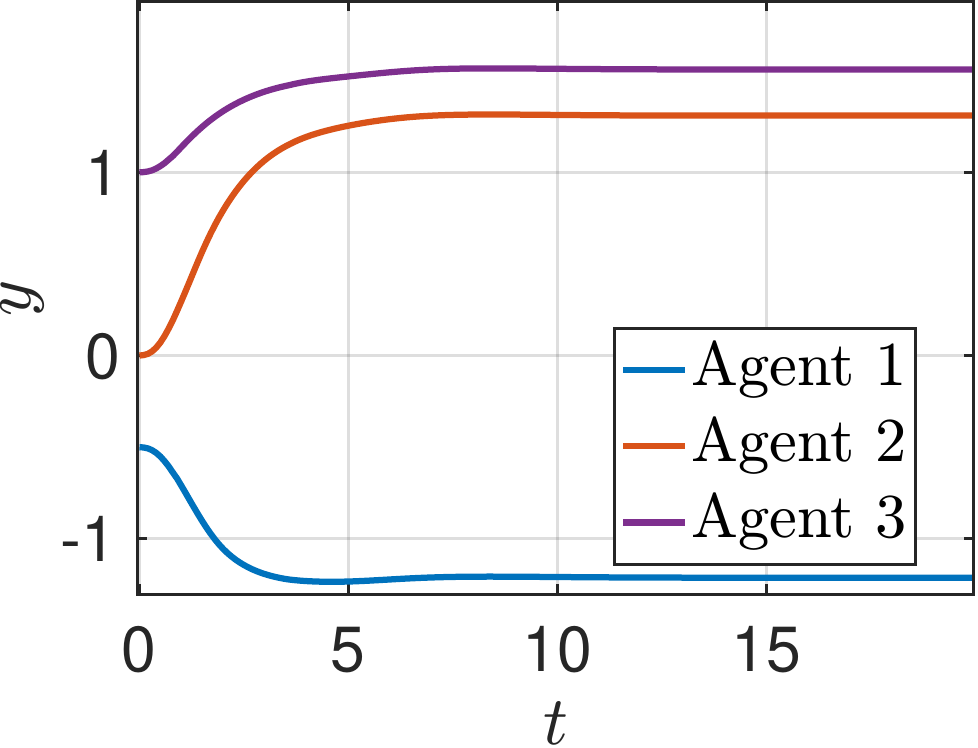}}
\subfigure[]{
		\includegraphics[scale=0.22]{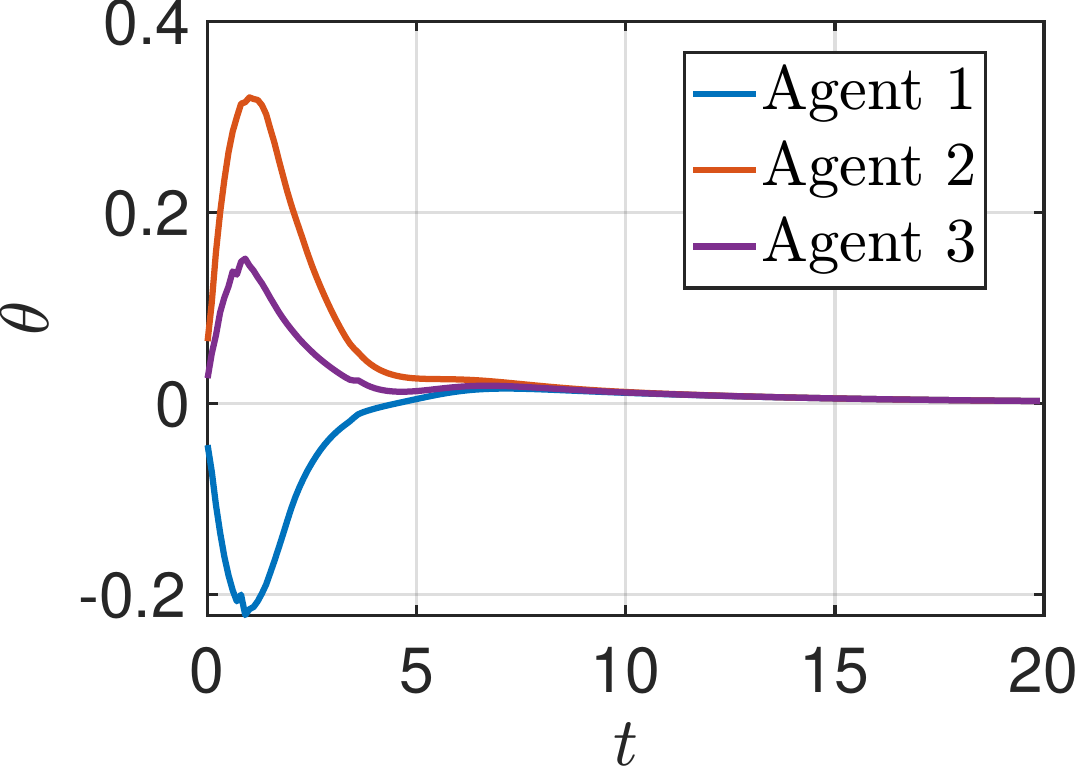}}
\subfigure[]{
        		\includegraphics[scale=0.22]{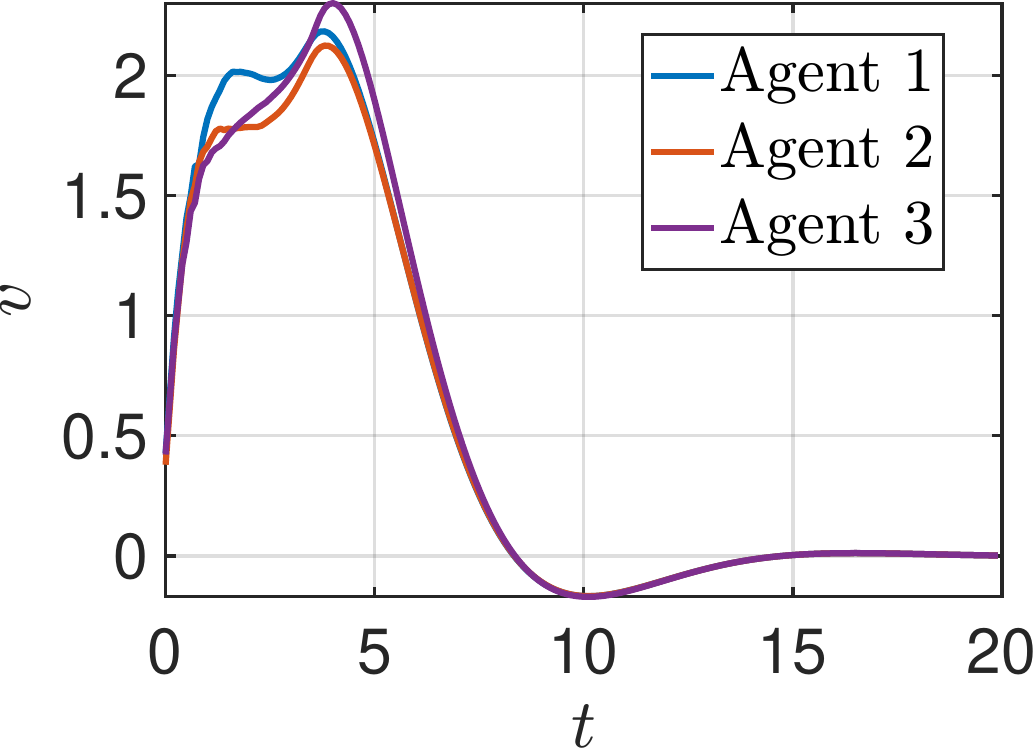}}
\caption{State trajectories of system \eqref{simdyn} under the CMPC.}
\label{xy}
\end{figure}

 \begin{figure}[t]
\centering
\subfigure[]{
		\includegraphics[scale=0.22]{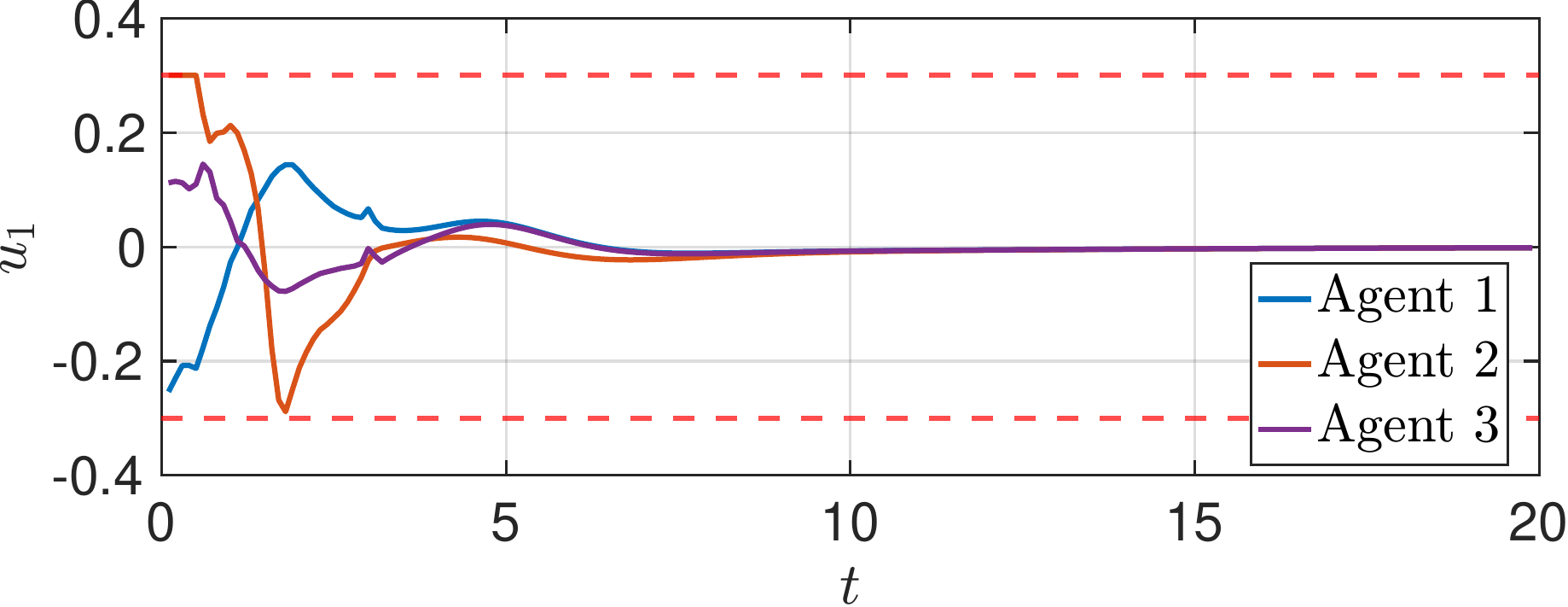}}
\subfigure[]{
		\includegraphics[scale=0.22]{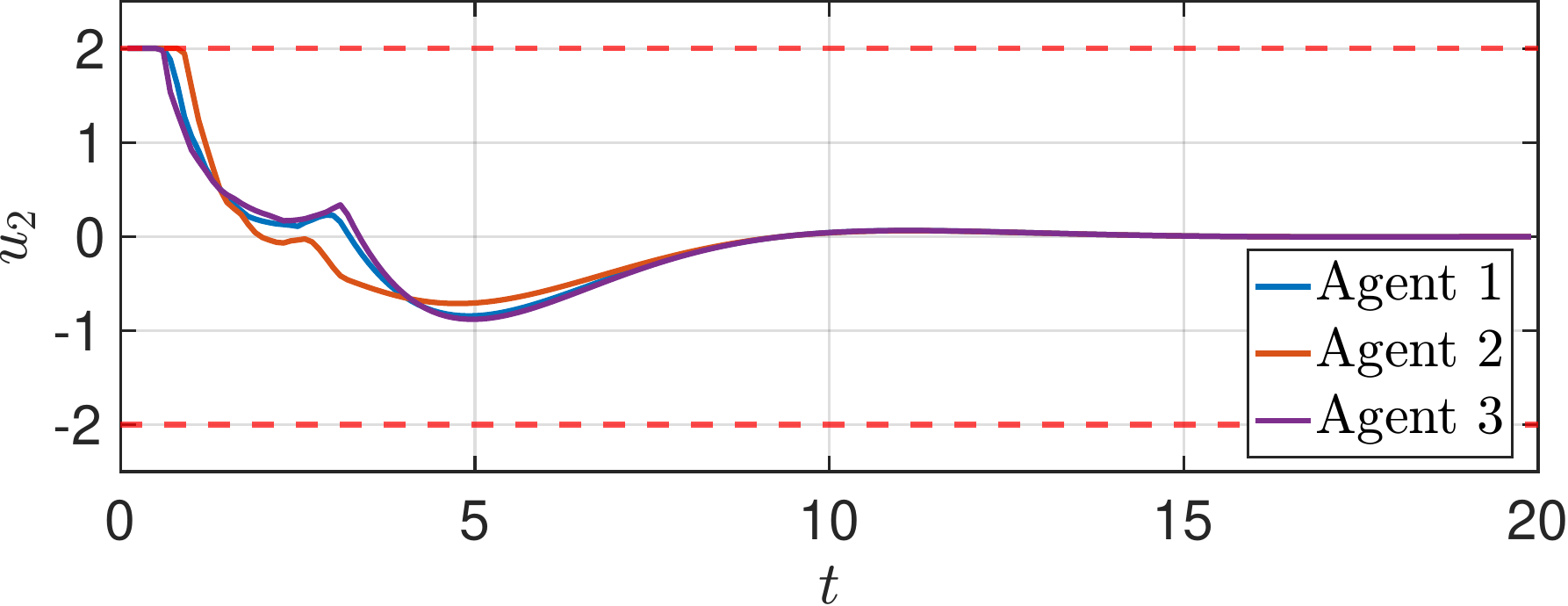}}
 \caption{Control inputs generated by the CMPC.}
 \label{u}
 \end{figure}
 
  \begin{figure}[t]
 \centering
     \includegraphics[scale=0.22]{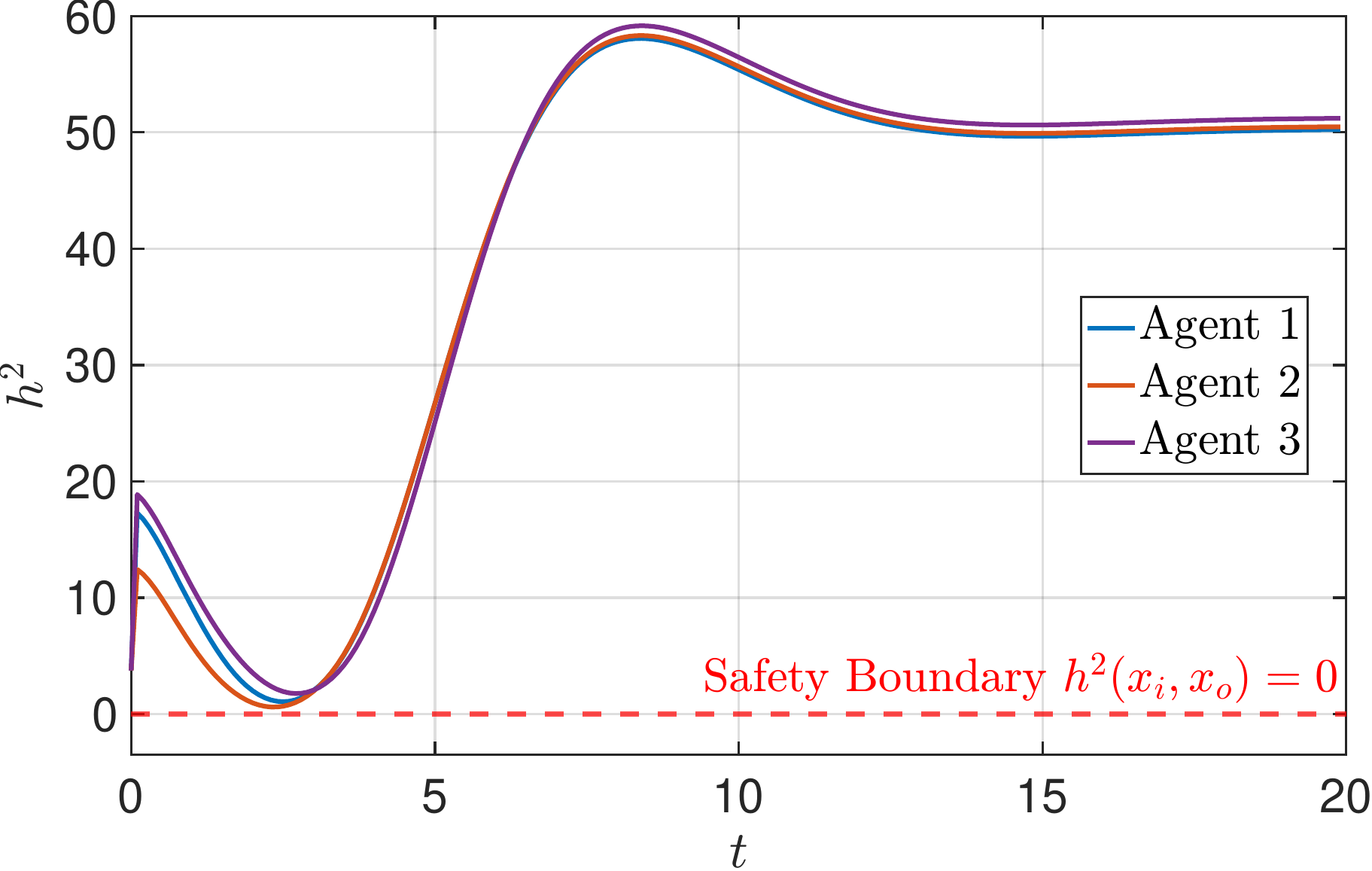}
 \caption{Time evolution of the DHCBF under the CMPC.}
 \label{h}
 \end{figure}
 
   \begin{figure}[t]
 \centering
     \includegraphics[scale=0.22]{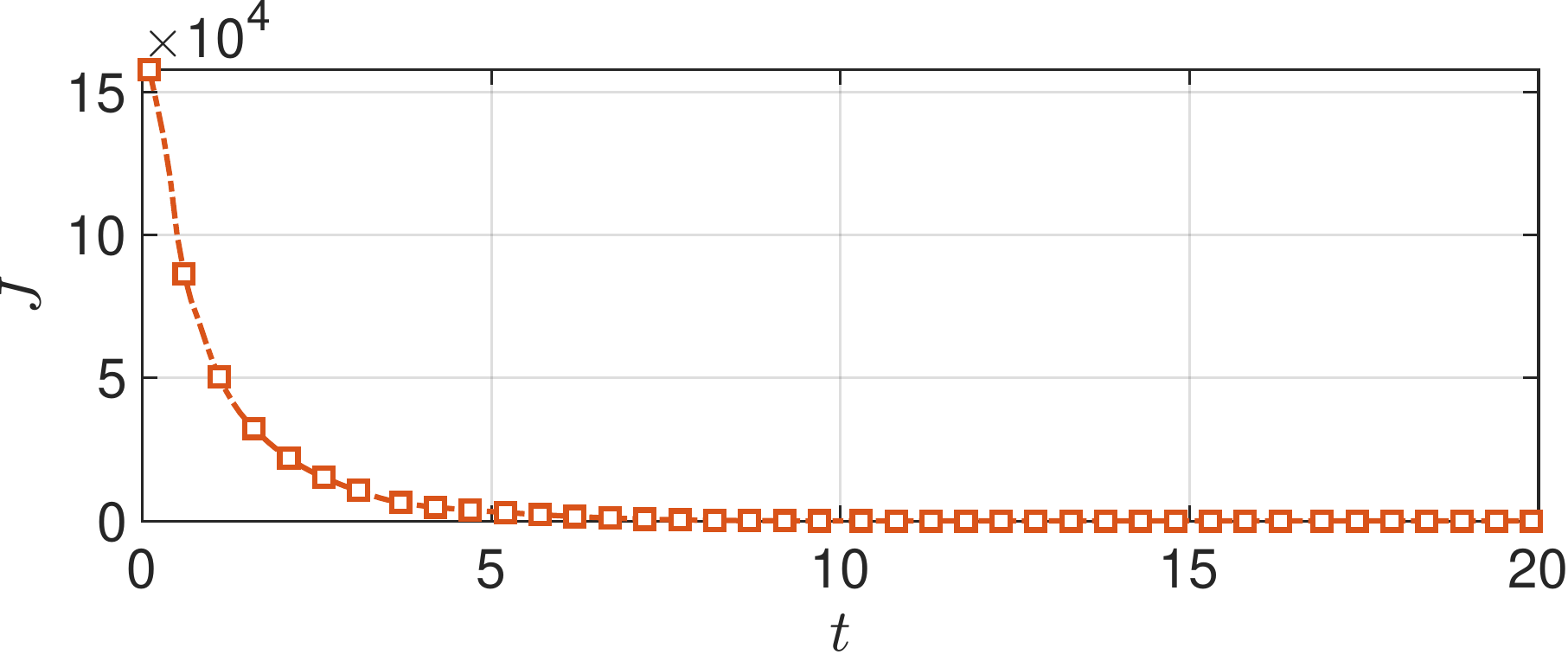}
 \caption{Time evolution of the cost function under the CMPC.}
 \label{J}
 \end{figure}
 
   \begin{figure}[t]
 \centering
     \includegraphics[scale=0.22]{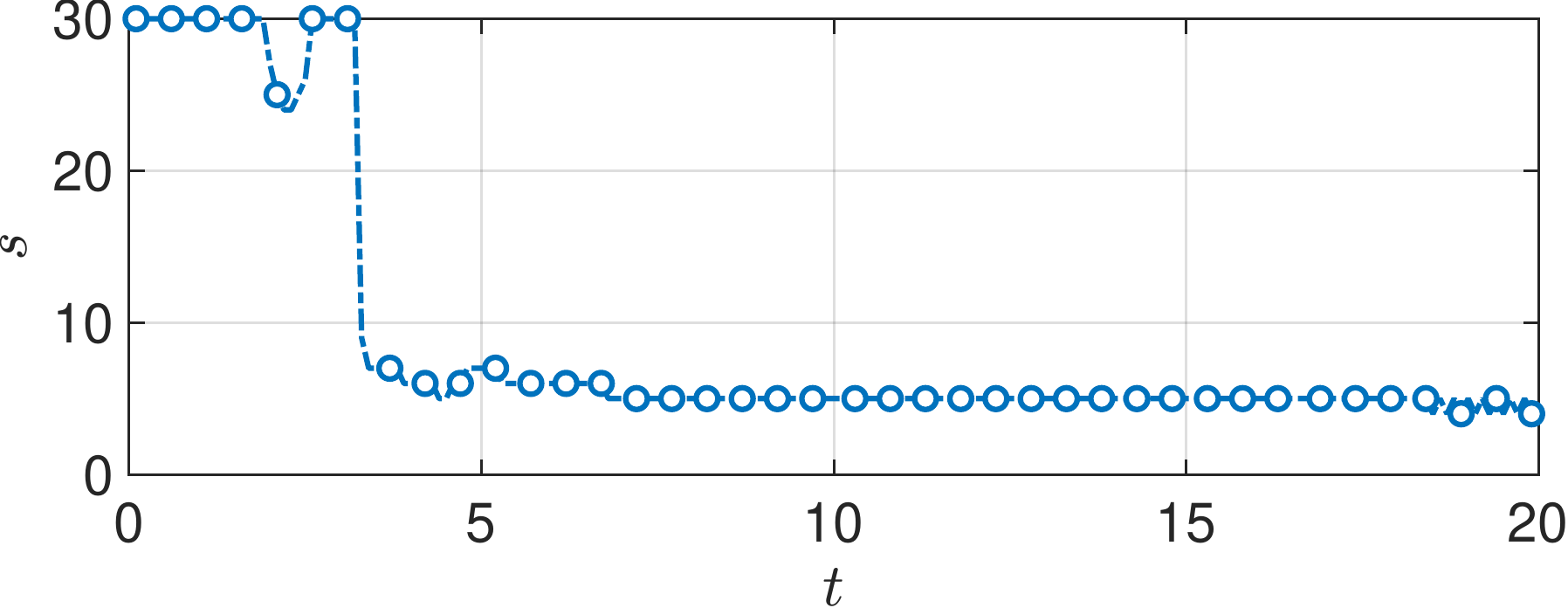}
 \caption{Time evolution of the iteration.}
 \label{s}
 \end{figure}

By using our proposed CMPC method, the simulation results validate the effectiveness and computational efficiency of the proposed CMPC approach. 
As illustrated in Fig. \ref{xy}, the output trajectories of all agents, including positions $p_{i,x}$, heading angles $\theta_i$, and linear velocities $v_i$, rapidly achieve consensus from distinct initial conditions. Fig. \ref{xy} (b) demonstrates that there is no collision between all agents, i.e., $h^{1}(x_{i},x_{j})>0$ for $t \geq 0$. Correspondingly, Fig. \ref{u} confirms that the control inputs $u_{i,1}$ and $u_{i,2}$ strictly remain within the prescribed constraints indicated by the red dashed lines throughout the simulation. The obstacle aovidance performance is shown in Fig. \ref{h}, where the values of the DHCBF $h^{2}(x_i, x_{o})$ for all agents are maintained strictly above the safety boundary $h^{2}(x_i, x_{o})=0$, ensuring that the system states remain within the safe set $\mathcal{C}$. Furthermore, the stability of CMPC is corroborated by Fig. \ref{J}, where the cost function value $J$ monotonically decreases and stabilizes at zero, which demonstrates Theorem \ref{stability}. Fig. \ref{s} demonstrates the efficiency of the SQP scheme. Initially ($t \le 3$), iterations hit the maximum limit $s_{\max}=30$ due to the poor quality of the zero-input guess. Subsequently, the iteration count drops and stabilizes at around 5. This is driven by the warm-start strategy, where the initial guesses for $t \ge 3$ become significantly closer to the local optimal solution, thereby facilitating faster convergence.

Finally, a quantitative comparison with other methods, including Distributed Safety-Critical MPC (DSMPC) in (\cite{wang2025distri}) and the MPC-Distance Constraints (MPC-DC) method in (\cite{RAVANSHADI202375}), is presented in Table \ref{tab1} regarding computational performance. The proposed method achieves an average computation time ($\mathrm{T_{\!avg}}$) of only $0.0371\,\text{s}$, which is significantly faster than the baselines MPC-DC ($1.291\,\text{s}$) and DSMPC ($1.952\,\text{s}$). Furthermore, the proposed CMPC demonstrates superior numerical stability with a maximum computation time ($\mathrm{T_{max}}$) limited to $0.1958\,\text{s}$, whereas DSMPC spikes to $11.712\,\text{s}$, indicating that the proposed CMPC method effectively avoids the heavy computational burden associated with solving non-convex NLP problems directly. Overall, our method achieves a speedup of approximately $35\times$ and $52\times$ compared to MPC-DC and DSMPC, respectively, in terms of average execution time. Furthermore, it reduces the total simulation time ($\mathrm{T_{total}}$) from hundreds of seconds to merely $7.42\,\text{s}$, validating its high efficiency and practicality for online applications.

\begin{table}[htbp]
    \centering
    \caption{Computation Time Comparison.}
    \label{tab1}
    \begin{tabular}{lccc}
        \toprule
        \textbf{Method} & $\mathbf{T_{avg}}$ \textbf{(s)} & $\mathbf{T_{max}}$ \textbf{(s)} & $\mathbf{T_{total}}$ \textbf{(s)} \\
        \midrule
        DSMPC     & 1.952          &  11.712         & 390.400       \\
        MPC-DC     & 1.291          & 7.359          & 258.220       \\
        \textbf{Ours} & \textbf{0.0371} & \textbf{0.1958} & \textbf{7.420} \\
        \bottomrule
    \end{tabular}
\end{table}

\section{Conclusion}
\label{6}
This paper has presented a novel CMPC approach to achieve safe output consensus for nonlinear multi-agent systems. By leveraging an SQP scheme with linearized dynamics and convexified DHCBF constraints, the proposed method effectively improves the computational efficiency. We have provided theoretical proofs for the local convergence of the SQP scheme, as well as the recursive feasibility and stability of the proposed CMPC. Quantitative comparisons in simulations confirmed that our approach reduces computation time compared to baseline methods, validating its potential for real-time applications. Future work will focus on extending this framework to a fully distributed implementation.

\bibliography{ref}

\end{document}